\documentclass[conference]{IEEEtran}
\IEEEoverridecommandlockouts
\usepackage{cite}
\usepackage{amsmath,amssymb,amsfonts}
\usepackage{algorithmic}
\usepackage{graphicx}
\usepackage{textcomp}
\usepackage{xcolor}
\usepackage[caption=false,font=footnotesize, subrefformat=parens]{subfig}
\usepackage{listings}
\usepackage{hyperref}

\hypersetup{colorlinks=true, urlcolor=black, linkcolor=black, citecolor=black}

\newcommand{\vect}[1]{{\mathbf{#1}}}
\newcommand{\matr}[1]{\mathbf{#1}}

\newcommand{\D}{\mathrm{d}}
\newcommand{\jj}{\mathrm{j}}
\newcommand{\JJ}{\matr{J}}
\newcommand{\T}{\mathrm{T}}                
\newcommand{\e}{\mathrm{e}}                
\newcommand{\I}{\matr{I}} 

\newcommand{\Ra}{R_{\mathrm{a}}}
\newcommand{\alphao}{\alpha_\mathrm{o}}
\newcommand{\alphapll}{\alpha_\mathrm{pll}}
\newcommand{\Ko}{\matr{K}_{\mathrm{o}}}
\newcommand{\kt}{k_{\mathrm{t}}}
\newcommand{\kpb}{\vect{k}_{\mathrm{p}}}
\newcommand{\kpbo}{\vect{k}_{\mathrm{p0}}}
\newcommand{\kp}{k_{\mathrm{p}}}
\newcommand{\kvb}{\vect{k}_{\mathrm{v}}}
\newcommand{\kvbo}{\vect{k}_{\mathrm{v0}}}
\newcommand{\kv}{k_{\mathrm{v}}}
\newcommand{\ki}{k_{\mathrm{i}}}
\newcommand{\alphac}{\alpha_\mathrm{c}}
\newcommand{\alphai}{\alpha_\mathrm{i}}
\newcommand{\ui}{\vect{u}_\mathrm{i}}

\newcommand{\omegag}{\omega_\mathrm{g}}
\newcommand{\hatomegag}{\hat{\omega}_\mathrm{g}}
\newcommand{\omegac}{\omega_\mathrm{c}}
\newcommand{\thetac}{\vartheta_\mathrm{c}}

\newcommand{\udc}{u_\mathrm{dc}}
\newcommand{\ug}{\vect{u}_\mathrm{g}}
\newcommand{\eg}{\vect{e}_\mathrm{g}}
\newcommand{\ugmago}{u_\mathrm{g0}}
\newcommand{\ig}{\vect{i}_\mathrm{g}}
\newcommand{\vc}{\vect{v}_\mathrm{c}}
\newcommand{\vcmag}{v_\mathrm{c}}
\newcommand{\hatvcmag}{\hat{v}_\mathrm{c}}
\newcommand{\hatvc}{\hat{\vect{v}}_\mathrm{c}}
\newcommand{\uc}{\vect{u}_\mathrm{c}}
\newcommand{\ucdo}{u_\mathrm{cd0}}
\newcommand{\ucqo}{u_\mathrm{cq0}}
\newcommand{\vcref}{v_\mathrm{c,ref}}
\newcommand{\ucref}{\vect{u}_\mathrm{c,ref}}

\newcommand{\qgref}{q_\mathrm{g,ref}}
\newcommand{\pgref}{p_\mathrm{g,ref}}
\newcommand{\pg}{p_\mathrm{g}}
\newcommand{\hatpg}{\hat{p}_\mathrm{g}}

\newcommand{\ugs}{\vect{u}_\mathrm{g}^\mathrm{s}}
\newcommand{\igs}{\vect{i}_\mathrm{g}^\mathrm{s}}
\newcommand{\egs}{\vect{e}_\mathrm{g}^\mathrm{s}}

\newcommand{\Lf}{L_\mathrm{f}}
\newcommand{\Lt}{L_\mathrm{t}}
\newcommand{\Lg}{L_\mathrm{g}}

\newcommand{\pgo}{p_\mathrm{g0}}
\newcommand{\qgo}{q_\mathrm{g0}}

\newcommand{\igref}{\vect{i}_\mathrm{g,ref}}
\newcommand{\Td}{T_\mathrm{d}}
\newcommand{\Ts}{{T_\mathrm{s}}}

\newcommand{\Yc}{\matr{Y}_\mathrm{c}}
\newcommand{\Ydd}{Y_\mathrm{dd}}
\newcommand{\Yqd}{Y_\mathrm{qd}}
\newcommand{\Ydq}{Y_\mathrm{dq}}
\newcommand{\Yqq}{Y_\mathrm{qq}}
\newcommand{\Zf}{\matr{Z}_\mathrm{f}}
\newcommand{\Zg}{\matr{Z}_\mathrm{g}}

\newcommand{\igd}{i_\mathrm{gd}}
\newcommand{\igdo}{i_\mathrm{gd0}}
\newcommand{\igqo}{i_\mathrm{gq0}}
\newcommand{\igo}{\vect{i}_\mathrm{g0}}

\newcommand{\hatvcmago}{\hat{v}_\mathrm{c0}}
\newcommand{\hatvco}{\hat{\vect{v}}_\mathrm{c0}}
\newcommand{\hatugmag}{\hat{u}_\mathrm{g}}
\newcommand{\hatLt}{\hat{L}_\mathrm{t}}

\newcommand{\Gtheta}{G_\vartheta}
\newcommand{\Gu}{\matr{G}_\mathrm{u}}
\newcommand{\Gi}{\matr{G}_\mathrm{i}}

\newcommand{\nuF}{\nu_\mathrm{F}}
\newcommand{\hatthetag}{\hat{\vartheta}_\mathrm{g}}
\newcommand{\tildethetag}{\tilde{\vartheta}_\mathrm{g}}
\newcommand{\ucmago}{u_\mathrm{c0}}

\newcommand{\ec}{\vect{e}_\mathrm{c}}
\newcommand{\ues}{\vect{u}_\mathrm{e}^\mathrm{s}}
\newcommand{\uff}{\vect{u}_\mathrm{ff}}

\newcommand{\omegae}{\omega_\mathrm{e}}
\newcommand{\fe}{f_\mathrm{e}}

\newcommand{\ugdprime}{u_\mathrm{gd}'}
\newcommand{\ugqprime}{u_\mathrm{gq}'}
\newcommand{\ugprime}{\mathbf{u}_\mathrm{g}'}
\newcommand{\igprime}{\mathbf{i}_\mathrm{g}'}

\newcommand{\Kone}{\mathbf{K}_\mathrm{1}}
\newcommand{\Ktwo}{\mathbf{K}_\mathrm{2}}

\definecolor{codegreen}{rgb}{0,0.6,0}
\definecolor{codepurple}{rgb}{0.3,0.3,0.3}

\lstdefinestyle{mystyle}{
    commentstyle=\color{codegreen},
    keywordstyle=\color{magenta},
    stringstyle=\color{codepurple},
    basicstyle=\ttfamily\footnotesize,
    breakatwhitespace=false,         
    breaklines=true,                 
    captionpos=b,                    
    keepspaces=true,                                
    showspaces=false,                
    showstringspaces=false,
    showtabs=false,                  
    tabsize=2
}
\newcounter{algorithm}
\renewcommand{\thealgorithm}{Algorithm~\arabic{algorithm}}

\begin{document}

\title{Open-Source Python Tool for Grid Converter Output Admittance Identification
\thanks{This project was supported by ABB Oy. The authors acknowledge the use of the EPE infrastructure of Aalto University School of Electrical Engineering.

\copyright 2026 IEEE. Personal use of this material is permitted. Permission from IEEE must be obtained for all other uses, in any current or future media, including reprinting/republishing this material for advertising or promotional purposes, creating new collective works, for resale or redistribution to servers or lists, or reuse of any copyrighted component of this work in other works.}
}

\author{\IEEEauthorblockN{Juho Määttä}
\IEEEauthorblockA{\textit{School of Electrical Engineering} \\
\textit{Aalto University}\\
Espoo, Finland \\
juho.k.maatta@aalto.fi}
\\
\IEEEauthorblockN{Jarno Kukkola}
\IEEEauthorblockA{\textit{ABB Oy} \\
\textit{Drives} \\
Helsinki, Finland \\
jarno.kukkola1@fi.abb.com}
\and
\IEEEauthorblockN{Janne Seppänen}
\IEEEauthorblockA{\textit{School of Electrical Engineering} \\
\textit{Aalto University}\\
Espoo, Finland \\
janne.seppanen@aalto.fi}
\\
\IEEEauthorblockN{Marko Hinkkanen}
\IEEEauthorblockA{\textit{School of Electrical Engineering} \\
\textit{Aalto University}\\
Espoo, Finland \\
marko.hinkkanen@aalto.fi}
}

\maketitle

\begin{abstract}
Frequency-domain analysis based on converter output admittance is a key tool for studying converter-driven stability in power grids. This paper presents a Python-based identification tool built on a completely open-source simulator, eliminating the need for commercial licenses such as MATLAB or PSCAD and improving configurability through an MIT-licensed stack. The identification method uses steady-state signal injection with a sinusoidal sweep, deriving frequency-domain admittance from time-domain simulations. Analytical output-admittance models are developed for both grid-forming (disturbance-observer-based) and grid-following (phase-locked-loop-based) control to verify the numerical results. The tool's results are compared against a commercial PSCAD-based alternative, demonstrating accurate admittance identification across control methods. Code and examples are available online to support reproducibility.
\end{abstract}


\section{Introduction}
Converter-based generation is rapidly increasing, and converter-driven stability has become a key concern, e.g., because converters can excite or amplify resonances in the grid~\cite{hatziargyriou2021}. To analyze such interactions, frequency-domain methods based on converter output admittance are widely used\cite{harnefors2007a,sun2011}. When the converter output admittance and grid impedance are known, stability can be assessed using the multivariable Nyquist criterion~\cite{skogestad}. Alternatively, passivity-based conditions can be applied using the converter output admittance alone~\cite{harnefors2016}.

The converter output admittance can be obtained either analytically, if the white-box model of the converter is available, or numerically from time-domain signals. Numerical identification can be performed either using simulation data or experimental measurements. The advantages of the numerical method are that tedious analytical calculations can be avoided and that the output admittance can also be obtained for a black-box converter model.

Several tools have been developed for numerical output-admittance identification and frequency-domain analysis in simulations. Commercial options include the AIM Toolbox~\cite{aim_toolbox2} and Grid Impedance Scan Tool~\cite{shah2022}, both relying on PSCAD as the simulation engine. Open-source projects have also been presented, such as Simplus Grid Tool~\cite{simplus} and Z-tool~\cite{cifuentes2024}. However, they still depend on commercial platforms, MATLAB and PSCAD, respectively. In some applications, the use of these commercial platforms may be beneficial due to their adoption in the industry. For example, Fingrid, the transmission system operator in Finland, requires PSCAD models to be submitted for certain power generating facilities and grid energy storage systems~\cite{vjv2024,sjv2024}. However, licensing costs and the limited configurability of proprietary software may still be issues.

To address these challenges, this paper presents a Python-based identification tool built on a completely open-source simulator platform, \textit{motulator}, published under the MIT license. The presented tool is available in the \textit{motulator} project at~\cite{motu}. Implementing the identification within the same Python project clarifies the data flow, simplifies configuration, and lowers barriers to adoption for both academia and industry.

The identification method uses steady-state signal injection with a sinusoidal frequency sweep to obtain the converter output admittance from time-domain simulations. A series voltage excitation is applied at the point of common coupling (PCC), and the grid-current response is measured. Two linearly independent injections ($d$- and $q$-axis) provide the four equations required to compute the 2$\times$2 output-admittance matrix via discrete Fourier transform (DFT) at each frequency. For each frequency, two simulation runs are performed and parallelized across CPU cores to accelerate the sweep.

The rest of this paper is structured as follows. Section~\ref{sec:modeling} presents the converter modeling and output admittance. Section~\ref{sec:numerical_tool} details the implementation of the identification tool. Section~\ref{sec:analytical_derivation} derives analytical output-admittance models for two control methods to verify the numerical results. Section~\ref{sec:results} compares the presented tool with a commercial identification tool and with the analytical models. Section~\ref{sec:conclusion} concludes the paper.

\section{Converter Output Admittance}
\label{sec:modeling}

\subsection{Large-Signal Model}

Fig.~\ref{fig:system_model}(a) shows the large-signal model of the system that is considered. The converter is connected to the grid through an output filter, which could be, e.g., an L or LCL filter. The grid is modeled as the ideal voltage source $\egs$ behind the impedance $\Zg^\mathrm{s}(s)$, which is assumed to comprise only series-connected inductance, capacitance, and resistance. The system is modeled in the time domain, and here $s=\D/\D t$ represents the time derivative operator. A three-phase system is considered, and space-vector notation with real column vectors is used. Furthermore, the superscript $\mathrm{s}$ denotes space vectors and transfer-function matrices in stationary coordinates.

The PCC is modeled between the grid impedance and the output filter. The PCC voltage is
\begin{align}
    \ugs = \Zg^\mathrm{s}(s)\igs + \ues + \egs
\end{align}
where $\igs$ is the grid current and $\ues$ is the injected excitation voltage. Fig.~\ref{fig:system_model}(a) also shows a generic controller, which typically uses measurements from the converter or grid current and the DC-bus voltage $\udc$. Additionally, depending on the control method, a PCC-voltage measurement may be used.

\subsection{Small-Signal Model}

\begin{figure}[!t]
    \centering
    \subfloat[]{\includegraphics[width=\linewidth]{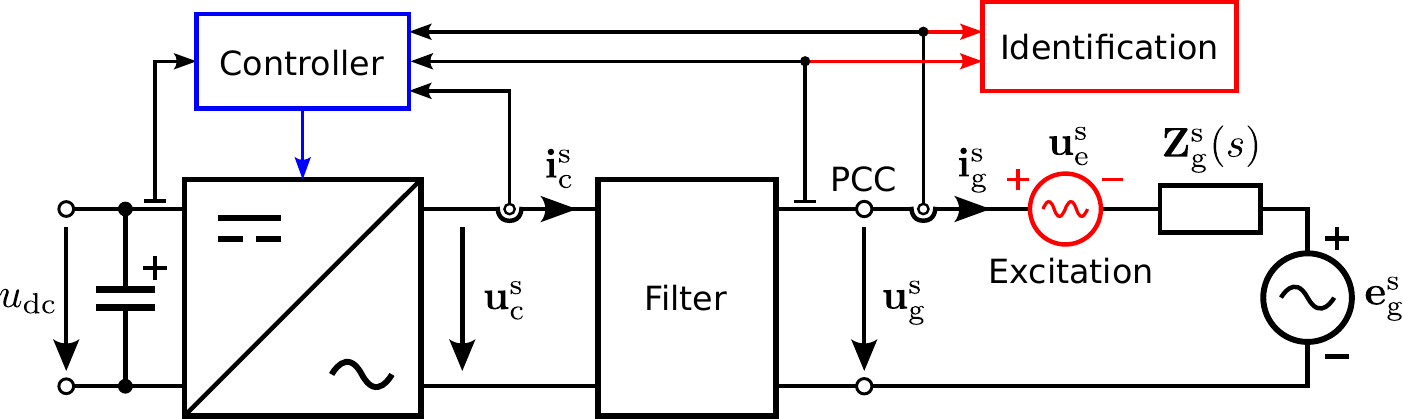}\label{fig:system_largesignal}}
    \hfil
    \subfloat[]{\includegraphics[width=0.75\linewidth]{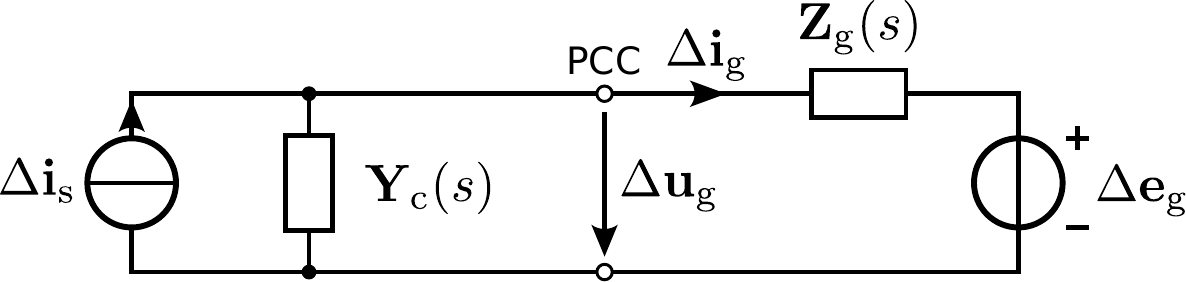}\label{fig:system_smallsignal}}
    \caption{System models: (a) large-signal model in stationary coordinates; (b) small-signal model in $dq$-coordinates. In (a), the parts added for numerical identification purposes are shown in red. Under normal operation, $\ues=0$. During signal injection, $\Zg^\mathrm{s}=0$ to decouple the converter from the grid.}
    \label{fig:system_model}
\end{figure}

For the purpose of frequency-domain stability analysis, the system is divided into separate subsystems for the converter and the grid~\cite{sun2011}. Fig.~\ref{fig:system_model}(b) shows the small-signal model obtained by linearizing the system in Fig.~\ref{fig:system_model}(a) around an operating point. The notation $\ig = \igo + \Delta\ig$ is used, where operating-point quantities are denoted by the subscript $0$ and small-signal deviations around the operating point by a preceding $\Delta$.

To define the steady-state operating point, the system is expressed in $dq$-coordinates rotating at the grid angular frequency $\omegag$, which is assumed to be constant. Same stability conclusions are reached if the coordinates are aligned with the PCC-voltage vector, incorporating deviations in the grid angular frequency into the angle of the synchronous coordinates~\cite{narula2025}. Space vectors and transfer-function matrices in $dq$-coordinates are shown without a superscript.

The converter, output filter, and controller are modeled as a Norton equivalent using the current source $\Delta\mathbf{i}_\mathrm{s}$ and the output admittance $\Yc$, which is defined as
\begin{align}
    \Yc (s) =
    \begin{bmatrix}
        \Ydd (s) & \Yqd (s)\\
        \Ydq (s) & \Yqq (s)
    \end{bmatrix}.
\end{align}

\section{Numerical Identification Tool}
\label{sec:numerical_tool}

\subsection{Identification Framework}
Common to all identification methods for black-box systems is injecting an excitation signal to the system and measuring the response. For the converter system, the excitation signal can be realized using series voltage excitation, as shown in Fig.~\ref{fig:system_model}(a). A signal is injected in series on the PCC, and the grid-current response is measured. Applying the DFT to the PCC voltage and grid current, the output admittance at the excitation frequency can then be calculated.

As the output admittance is a 2$\times$2 matrix, four linearly independent equations are required to calculate the four unknown elements. This can be realized by using two linearly independent signal injections, and measuring the $d$- and $q$-axis variables for both injections. The output admittance is calculated as~\cite{francis2011}
\begin{align}
    \Yc(\jj\omega)= -
    &\begin{bmatrix}
        i_\mathrm{gd1}(\omegae) & i_\mathrm{gd2}(\omegae)\\
        i_\mathrm{gq1}(\omegae) & i_\mathrm{gq2}(\omegae)
    \end{bmatrix}\nonumber\\
    &\cdot
    \begin{bmatrix}
        u_\mathrm{gd1}(\omegae) & u_\mathrm{gd2}(\omegae)\\
        u_\mathrm{gq1}(\omegae) & u_\mathrm{gq2}(\omegae)
    \end{bmatrix}^{-1} \label{eq:admittance_numerical}
\end{align}
where the subscripts $1$ and $2$ refer to the two independent injections.

\subsubsection{Excitation Signal}

A sinusoidal excitation signal is used since this provides the most accurate results~\cite{roinila2013} and the operating point changing due to a long measurement time is not an issue in simulations. The excitation voltage in stationary coordinates is
\begin{equation}
  \mathbf{u}_\mathrm{e1}^\mathrm{s} = \e^{\omegag t\JJ}
  \begin{bmatrix}
    u_\mathrm{e}\\
    0
  \end{bmatrix}
  \cos(\omegae t)
\end{equation}
for the first injection in the $d$-axis, where $\JJ=\left[\begin{smallmatrix}0&-1\\1&0\end{smallmatrix}\right]$ is the orthogonal rotation matrix. The magnitude and angular frequency of the excitation voltage are $u_\mathrm{e}$ and $\omegae$, respectively. The excitation voltage is defined similarly for the second injection in the $q$-axis.

\begin{figure}[t]
    \centering
    \includegraphics[width=0.6\linewidth]{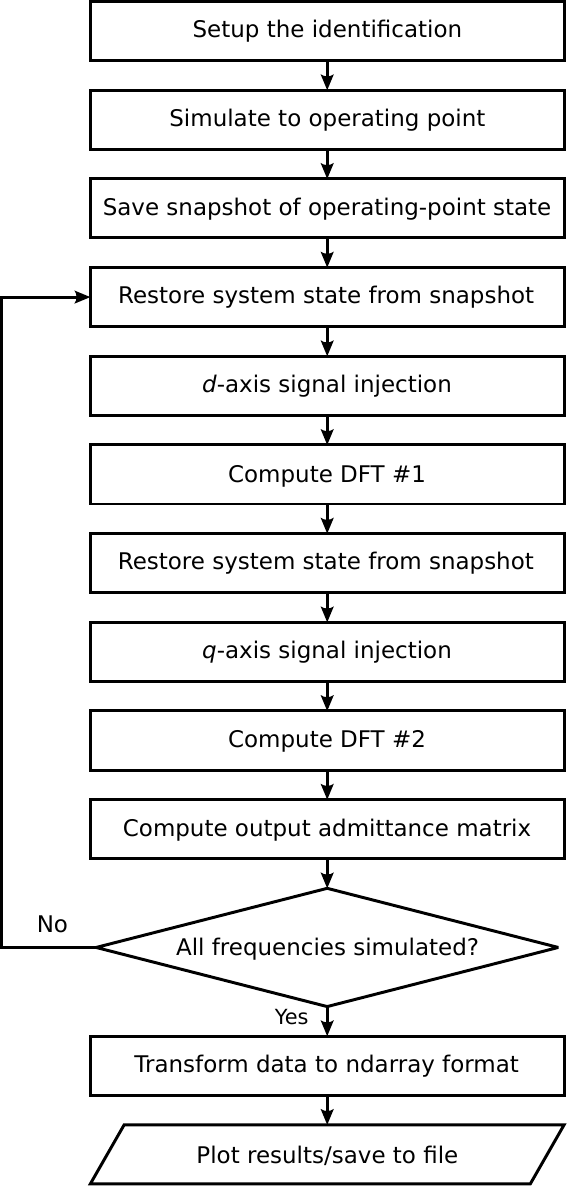}
    \caption{Flowchart showing the numerical identification procedure.}
    \label{fig:identification_flowchart}
\end{figure}

\subsubsection{Discrete Fourier Transform}

The excitation-frequency components of the measured signals are calculated using the DFT algorithm~\cite{proakis}. For instance, the DFT for the $d$-axis grid current is
\begin{equation}\label{eq:dft}
  \igd(\omegae) = \frac{2}{N}\sum_{n=0}^{N-1}\igd(n)W(n) \e^{-\jj 2\pi kn/N}
\end{equation}
where $N$ the number of samples, $n$ is the sample index, and $W(n)$ is a window function. The frequency index is $k = \fe N\Ts$, where $T_s$ is the sampling period of the measurements.

The result $\igd(\omegae)$ from the DFT is an excitation-frequency phasor, where the scaling factor $2/N$ is used for peak-value scaling. In total, eight Fourier transforms are performed at each excitation frequency to obtain the eight phasors required for calculating the admittance using~\eqref{eq:admittance_numerical}.

\subsection{Implementation Aspects}

The excitation signal is implemented as a sinusoidal sweep with logarithmically spaced frequencies, where the start and stop frequencies as well as the number of measurement frequencies can be configured. Optionally, the frequencies can be spaced linearly or a vector containing desired frequencies can be directly given.

As the converter output filter is dominantly inductive, the excitation-voltage magnitude is increased with the excitation frequency to allow for large enough current magnitude at higher frequencies. At the start excitation frequency, the magnitude specified by the user is used. Then, the magnitude is increased linearly until the stop frequency, by a factor set with a configuration parameter. The required voltage magnitude depends on the total inductance and the excitation frequency, and should be configured to keep the current magnitude under 0.1~p.u. over the whole frequency range to keep the small-signal assumption valid~\cite{riccobono2018}.

The use of the window function $W(n)$ in~\eqref{eq:dft} further reduces noise from the results at higher excitation frequencies. The Blackman window from the SciPy library is used~\cite{virtanen2020}, and use of the window can be disabled via a configuration parameter.

\subsection{Identification Procedure}

Fig.~\ref{fig:identification_flowchart} shows a flowchart of the identification process. After the identification parameters and simulation model have been configured, the system is simulated to reach the desired operating point. A snapshot of the operating-point state is taken, and the following identification runs are started from the snapshot. During signal injection, the grid impedance is set to zero and the grid-voltage magnitude is set to the magnitude of the operating-point PCC voltage. This ensures that the operating point, which is affected by the grid impedance, is retained but the converter output admittance is accurately identified without the effect of the grid.

Two simulation runs are required for the two signal injections at each excitation frequency. To accelerate identification, multiple simulations are run in parallel using Python's multiprocessing library~\cite{python-mp}. The operating point is first simulated in a single process and then the identification runs are performed in parallel on all available CPU cores.

\begin{figure*}[t]
    \centering
    \includegraphics[width=0.8\linewidth]{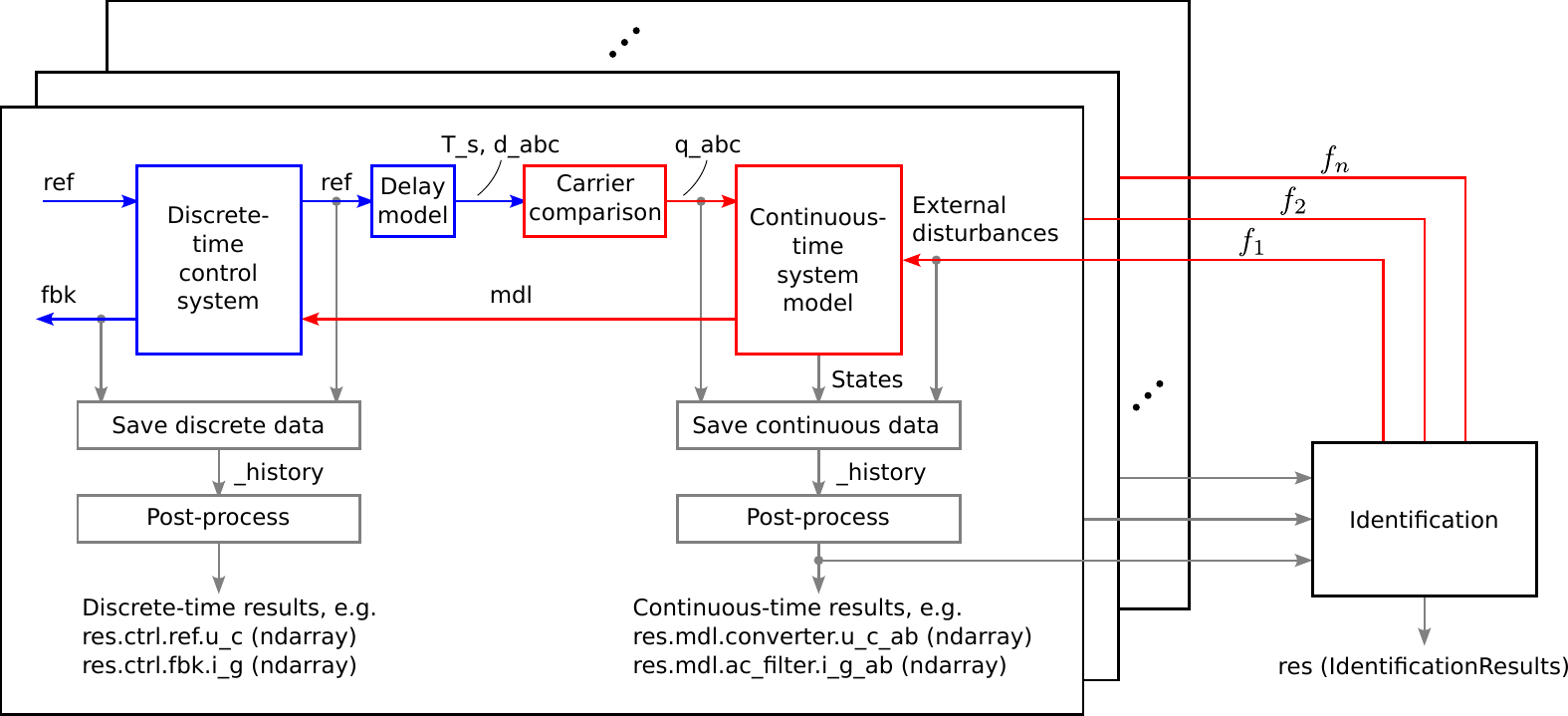}
    \caption{Block diagram illustrating the structure of the simulation software and the identification method.}
    \label{fig:software_structure}
\end{figure*}

Fig.~\ref{fig:software_structure} shows the data flow in the program and the interface between the system model and the identification method. After the identification is complete, the results and operating-point quantities are saved in an IdentificationResults object in the NumPy array format~\cite{harris2020}. By default, the results are then plotted. Optionally, results can also be saved to a file for later use.

\begin{figure}[t]
    \hrule
    \vspace{4pt}
    \refstepcounter{algorithm}
    \footnotesize \thealgorithm. Example script showing how the presented identification method can be used with the default configuration.
    \label{alg:code_example}
    \vspace{4pt}
    \hrule
    \centering
\begin{lstlisting}[language=Python, basicstyle=\scriptsize\ttfamily]
from motulator.grid import control, model, utils
from motulator.grid.utils._identification import (
    IdentificationCfg,
    plot_identification,
    run_identification,
)

# Compute base values based on the nominal values.
nom = utils.NominalValues(U=400, I=18, f=50, P=12.5e3)
base = utils.BaseValues.from_nominal(nom)

# Configure the identification.
identification_cfg = IdentificationCfg(
    abs_u_e=0.01*base.u
)

# Configure the system model.
ac_filter = model.LFilter(L_f=0.15*base.L)
ac_source = model.ThreePhaseSource(
    w_g=base.w, e_g=base.u
)
converter = model.VoltageSourceConverter(u_dc=650)
mdl = model.GridConverterSystem(
    converter, ac_filter, ac_source
)

# Configure the control system and set references.
inner_ctrl = control.ObserverBasedGridFormingController(
    i_max=1.3*base.i,
    L=0.15*base.L,
    T_s=identification_cfg.T_s,
)
ctrl = control.GridConverterControlSystem(inner_ctrl)
ctrl.set_power_ref(0.5 * base.p)
ctrl.set_ac_voltage_ref(base.u)

# Run the identification and plot results.
res = run_identification(
    identification_cfg, mdl, ctrl
)
plot_identification(res)
\end{lstlisting}
\hrule
\end{figure}

\subsection{Usage}

To use the identification method, the user needs only to run a Python script which defines the system model, controller, and identification settings. One example of such a script is given in Algorithm 1, which shows how the user can run the identification method with the minimum required configuration. By default, only the magnitude of the excitation voltage needs to be specified for the identification. The configurations of the system model and controller are identical to the other grid converter examples included in \textit{motulator}.

The identification can be extensively configured by setting the parameters of the IdentificationCfg data class. The system model and control system can also be configured. For example, only one line needs to be changed to use an LCL filter instead of the default L filter. Other modifications are also straightforward, such as using a different converter model or control method.

\section{Analytical Output Admittance}
\label{sec:analytical_derivation}

In this section, the analytical derivation of the converter output admittance is presented to verify the numerical results. As examples, the output-admittance calculation is presented for a disturbance-observer-based grid-forming (DO-GFM) control method~\cite{nurminen2024} and a grid-following (GFL) control method.

\subsection{Analytical Framework}
The goal in the analytical derivation of the converter output admittance is obtaining the linearized transfer-function matrix $\Yc(s)$ corresponding to
\begin{align}\label{eq:output_admittance}
    \Delta\ig = -\Yc(s)\Delta\ug.
\end{align}
Considering the system shown in Fig.~\ref{fig:system_model}(a), the analysis is separated into the linear system model and (typically) nonlinear controller. As an example, a system model with an L filter is considered, i.e., the grid-current dynamics are
\begin{align}\label{eq:ig_dynamics}
    \frac{\D\ig}{\D t} = \frac{1}{\Lf}\left(\uc - \ug - \omegag\Lf\JJ\ig\right).
\end{align}

The simulation tool implements the control algorithms in discrete time, but here the control system is represented in continuous time for the analytical output-admittance derivation. The control system can be represented as
\begin{subequations}\label{eq:ctrl_ss}
\begin{align}
    \frac{\D\mathbf{x}_\mathrm{c}}{\D t} &= \mathbf{f}_\mathrm{c}\left(\mathbf{x}_\mathrm{c},\ \mathbf{y},\ \mathbf{r}\right)\\
    \uc &= \mathbf{h}_\mathrm{c}\left(\mathbf{x}_\mathrm{c},\ \mathbf{y},\ \mathbf{r}\right)
\end{align}
\end{subequations}
where $\mathbf{x}_\mathrm{c}$ contains the controller states, $\mathbf{y}$ the feedback signals, and $\mathbf{r}$ the external references.

Assuming a constant DC-bus voltage, the feedback signals from the system model to the controller are $\mathbf{y} = \left[\ig^\T,\ \ug^\T\right]^\T$. The effect of pulse-width modulation is not included, and the converter output voltage is assumed to follow its reference value through a zero-order hold (ZOH) in stationary coordinates. A one-and-half-sampling-period time delay is used to model the ZOH and computational delay. Perfect compensation is assumed for the rotation of the synchronous coordinates due to the delays, leaving only the time delay modeled as
\begin{align}
    \uc = \e^{-s\Td}\ucref
\end{align}
where $\Td=1.5\Ts$. By linearizing $\mathbf{f}_\mathrm{c}$ and $\mathbf{h}_\mathrm{c}$, including the time delay, and assuming $\Delta\mathbf{r} = 0$, the converter voltage can be expressed as
\begin{align}\label{eq:uc_linearized}
    \Delta\uc = \e^{-s\Td}\left[\Gu(s)\Delta\ug - \Gi(s)\Delta\ig\right].
\end{align}

Equation~\eqref{eq:ig_dynamics} can be written in the Laplace domain and using small-signal notation as
\begin{subequations}\label{eq:ig}
\begin{align}\label{eq:ig_smallsignal}
    \Delta\ig = \Zf^{-1}(s)\left(\Delta\uc - \Delta\ug\right)
\end{align}
where
\begin{align}
    \Zf(s) = \Lf\left(s\I + \omegag\JJ\right).
\end{align}
\end{subequations}
Substituting~\eqref{eq:uc_linearized} into~\eqref{eq:ig_smallsignal}, and solving for $\Delta\ig$, then yields an equation corresponding to~\eqref{eq:output_admittance}, from which the output admittance can be extracted.

\subsection{Grid-Forming Control}

\subsubsection{Control Method Description}\label{sec:do-gfm_description}

\begin{figure}[!t]
    \centering
    \subfloat[]{\includegraphics[width=0.9\linewidth]{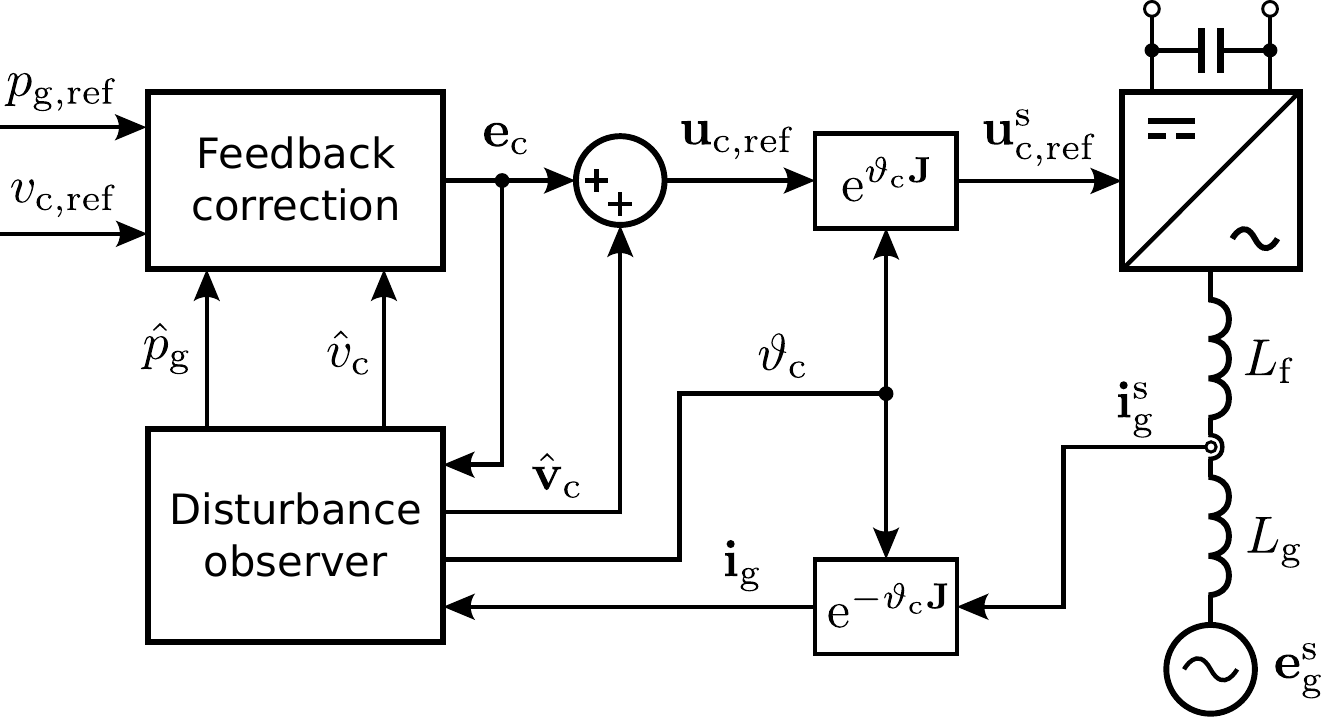}\label{fig:observer_blockdiagram}}
    \hfil
    \subfloat[]{\includegraphics[width=\linewidth]{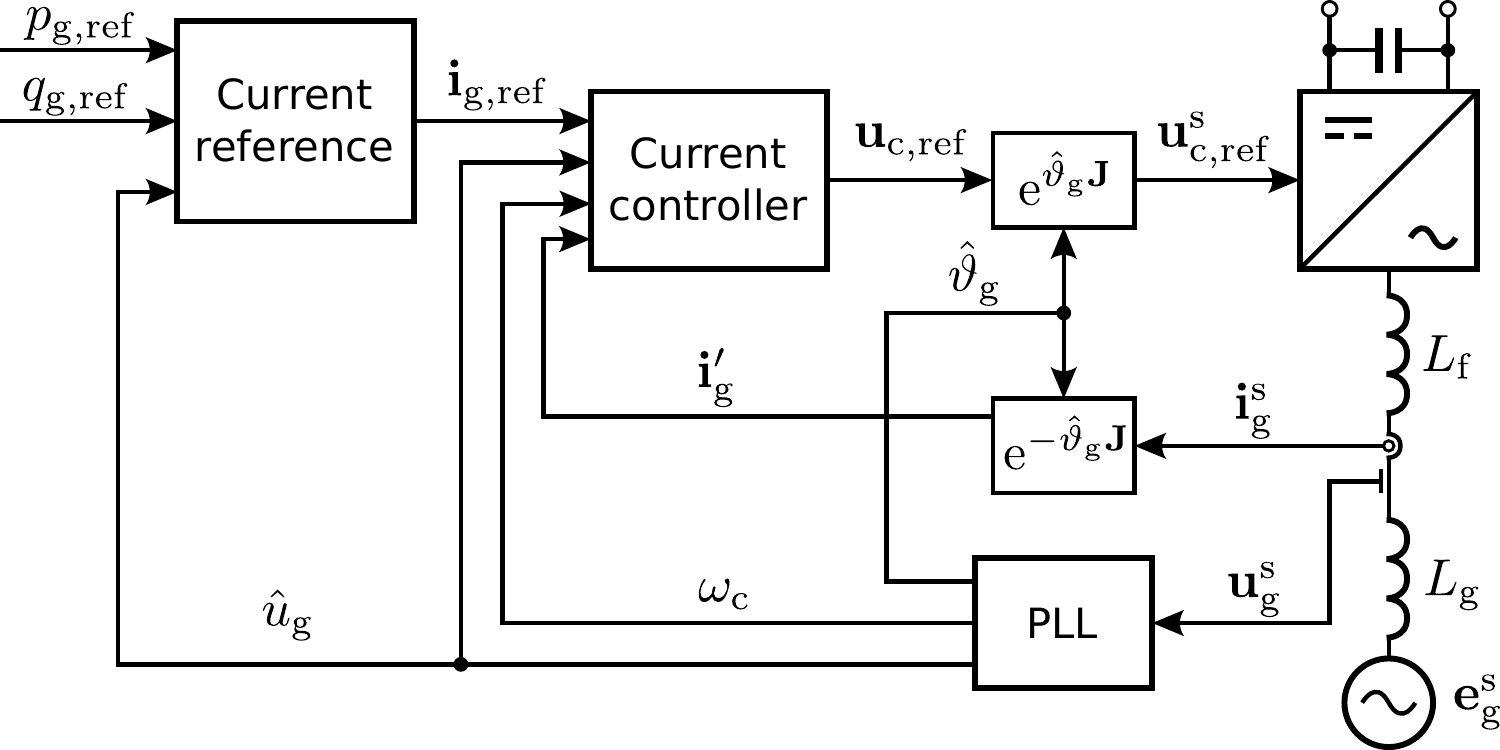}\label{fig:gfl_blockdiagram}}
    \caption{Example control methods: (a) disturbance-observer-based grid-forming control; (b) grid-following control.} 
    \label{fig:control_methods}
\end{figure}

Fig.~\ref{fig:control_methods}(a) shows a block diagram of the studied DO-GFM control method, which comprises a disturbance observer and a control law incorporated into the feedback-correction term~\cite{nurminen2024}. The estimate $\hatvc$ of the quasi-static converter voltage $\vc = \eg + \omegag\Lt\JJ\ig$, which in steady-state equals the converter output voltage $\uc$, is used as a state variable. The control method can be parametrized to have practically identical performance as reference-feedforward power-synchronization control (RFPSC)~\cite{harnefors2020b}.

The control system is implemented in coordinates rotating at the constant angular frequency $\D\thetac / \D t = \hatomegag$, and in these coordinates the disturbance observer is
\begin{subequations}\label{eq:observer}
\begin{align}
    \frac{\D\hatvc}{\D t} &= \left(\hatomegag\JJ + \Ko\right)\ec - \hatLt\Ko\frac{\D\ig}{\D t}\\
    \hatpg &= \frac{3}{2}\hatvc^\T \ig\\
    \hatvcmag &= ||\hatvc||
\end{align}
\end{subequations}
where $\Ko$ the observer gain matrix, $\ec$ the feedback-correction term, and $\hatLt$ the estimate for the total inductance $\Lf + \Lg$. In grid-forming mode, the active power $\pg$ and quasi-static converter-voltage magnitude $\vcmag$ are directly controlled, and estimates for these quantities are obtained from the disturbance observer. The feedback-correction term and converter output-voltage reference are
\begin{subequations}\label{eq:ec}
\begin{align}
    \ec &= \kpb(\pgref - \hatpg) + \kvb(\vcref - \hatvcmag)\\
    \ucref &= \ec + \hatvc
\end{align}
\end{subequations}
where $\kpb$ and $\kvb$ are the power-control and voltage-control gains, respectively. The control gains are selected as
\begin{align}
    \kpb = \frac{2\Ra}{3\vcref}\frac{\hatvc}{\hatvcmag} \quad \kvb = (\I - \kv\JJ)\frac{\hatvc}{\hatvcmag} \quad \Ko &= \alphao\I - \hatomegag\JJ
\end{align}
which correspond to the RFPSC-style gain selection in~\cite{nurminen2024}.

\subsubsection{Output Admittance}\label{sec:do-gfm_admittance}
The state variables of the control system are $\mathbf{x}_\mathrm{c} = \hatvc$. Using the linearized forms of \eqref{eq:observer} and \eqref{eq:ec} yields an expression for the converter output voltage corresponding to \eqref{eq:uc_linearized}, where
\begin{subequations}
\begin{align}\label{eq:do-gfm_linearized}
    \Gu(s) &= \mathbf{0}\\
    \Gi(s) &= \left(\I -\Kone\right)\left[s\I + \left(\hatomegag\JJ + \Ko\right)\Kone\right]^{-1}\nonumber\\
    &\quad \cdot \left[\left(\hatomegag\JJ + \Ko\right)\Ktwo + s\hatLt\Ko\right] + \Ktwo\\
    \Kone &= \frac{3}{2}\kpbo\igo^\T + \frac{1}{\hatvcmago}\kvbo\hatvco^\T\\
	\Ktwo &= \frac{3}{2}\kpbo\hatvco^\T.
\end{align}
\end{subequations}
Applying \eqref{eq:ig}, the resulting converter output admittance is.
\begin{align}
    \Yc(s) = \left[\Zf(s) + \e^{-s\Td}\Gi(s)\right]^{-1}.
\end{align}

\subsection{Grid-Following Control}

\subsubsection{Control Method Description}
Fig.~\ref{fig:control_methods}(b) shows a block diagram of the studied GFL control method. A typical PLL is used for tracking the PCC-voltage vector\cite{kaura1997}, with low-pass filtering of the PCC-voltage magnitude. The estimation dynamics for the PCC-voltage magnitude, PCC-voltage angle, and grid angular frequency, respectively, are
\begin{subequations}\label{eq:pll_dynamics}
\begin{align}
    \frac{\D\hatugmag}{\D t} &= 2\alphapll\left(\ugdprime - \hatugmag\right)\\
    \frac{\D\hatthetag}{\D t} &= \hatomegag + \frac{2\alphapll}{\hatugmag}\ugqprime = \omegac \label{eq:pcc_angle}\\
    \frac{\D\hatomegag}{\D t} &= \frac{\alphapll^2}{\hatugmag}\ugqprime
\end{align}
\end{subequations}
where $\alphapll$ is the frequency-tracking bandwidth of the PLL. The measured PCC voltage is transformed to $dq$-coordinates using the estimated PCC-voltage angle as $\ugprime = \exp(-\hatthetag\JJ)\ugs$.

A two-degree-of-freedom proportional-integral current controller with PCC-voltage feedforward is considered. The state-space form of the current controller is
\begin{subequations}\label{eq:current_ctrl}
\begin{align}
    \frac{\D\ui}{\D t} &= \left(\ki\I + \omegac\kt\JJ\right)\left(\igref - \igprime\right)\\
    \ucref &= \kt\igref - \kp\igprime + \ui + \uff
\end{align}
\end{subequations}
where $\uff = \left[\hatugmag,\ 0\right]^\T$ is the feedforward voltage and $\ki$, $\kp$, and $\kt$ are the integral, proportional, and feedforward gains, respectively. The gains can be selected as
\begin{align}
    \ki = \alphac\alphai\hatLt \quad \kp = \left(\alphac + \alphai\right)\hatLt \quad \kt = \alphac\hatLt
\end{align}
where $\alphac$ and $\alphai$ are the reference-tracking and integral-action bandwidths of the current controller, respectively. The grid current in estimated PCC-voltage coordinates is $\igprime = \exp(-\hatthetag\JJ)\igs$. Furthermore, the current reference is calculated as
\begin{align}
    \igref = \frac{2}{3\hatugmag}
    \begin{bmatrix}
        \pgref\\ -\qgref
    \end{bmatrix}.
\end{align}

\subsubsection{Output Admittance}\label{sec:gfl_admittance}

The linearization requires a steady-state operating point where time derivatives are zero. Therefore, instead of the PCC-voltage-angle estimation dynamics from~\eqref{eq:pcc_angle}, the estimation-error dynamics
\begin{align}\label{eq:thetag_est_error}
    \frac{\D\tildethetag}{\D t} = \omegag - \hatomegag - \frac{2\alphapll}{\hatugmag}\ugqprime
\end{align}
are considered. The controller states are then
\begin{align}
    \mathbf{x}_\mathrm{c} = \left[\hatugmag,\ \tildethetag,\ \hatomegag,\ u_\mathrm{id},\ u_\mathrm{iq}\right]^\T.
\end{align}
Combining the linearized forms of \eqref{eq:pll_dynamics}--\eqref{eq:thetag_est_error} yields the expression for the converter output voltage corresponding to~\eqref{eq:uc_linearized}, where
\begin{subequations}
\begin{align}
    \Gu(s) &=
    \begin{bmatrix}
        G_\mathrm{u11}(s) & G_\mathrm{u12}(s)\\
        G_\mathrm{u21}(s) & G_\mathrm{u22}(s)
    \end{bmatrix}\\
    G_\mathrm{u11}(s) &= \frac{2\alphapll}{s + 2\alphapll}\left[1 - \frac{1}{\ugmago}\left(\ki\igdo - \kt\igqo\right)\right]\\
    G_\mathrm{u12}(s) &= -\Gtheta(s)\left[\frac{\kt}{s}\omegag\igdo + \left(\kp + \frac{\ki}{s}\right)\igqo + \ucqo\right]\\
    G_\mathrm{u21}(s) &= -\frac{1}{\ugmago}\frac{2\alphapll}{s + 2\alphapll}\left(\kt\omegag\igdo + \ki\igqo\right)\\
    G_\mathrm{u22}(s) &= \Gtheta(s)\left[\left(\kp + \frac{\ki}{s}\right)\igdo - \frac{\kt}{s}\omegag\igqo + \ucdo\right]\\
    \Gtheta(s) &= \frac{1}{\ugmago}\frac{2\alphapll s + \alphapll^2}{s^2 + 2\alphapll s + \alphapll^2}\\
    \Gi(s) &= \left(\kp + \frac{\ki}{s}\right)\I + \frac{\kt}{s}\omegag\JJ.
\end{align}
\end{subequations}
Applying \eqref{eq:ig}, the resulting converter output admittance is
\begin{align}
    \Yc(s) = \left[\Zf(s) + \e^{-s\Td}\Gi(s)\right]^{-1}\left[\I - \e^{-s\Td}\Gu(s)\right].
\end{align}

\section{Results}
\label{sec:results}

In this section, results from the presented identification tool are compared against a commercial identification tool and the analytical solutions derived in Section~\ref{sec:analytical_derivation}. Further application of the results for calculation of the passivity index is also demonstrated.

\subsection{Identification Setup}

The frequency sweep is configured to measure one hundred logarithmically spaced frequencies from 1 Hz to 10 kHz. Other parameters for the example cases are listed in Table~\ref{tab:parameters}. The excitation-voltage magnitude is 0.01 p.u. at the start frequency, and is linearly increased to 0.05 p.u. at the stop frequency. In \textit{motulator}, the default variable-step solver is used and the solver is configured to return ten data points for each controller sampling period, corresponding to a 10 µs step size for the signals used in the identification. In the PSCAD model used for the commercial identification tool, a fixed step size of 20 µs is used.

To simplify the derivation of analytical models, the control methods used in this paper differ slightly from those provided in the main \textit{motulator} GitHub repository. An archived version is available in~\cite{github_archive} to reproduce the exact results presented in this paper. Along with the presented numerical identification tool, scripts for plotting the analytical solutions presented in this paper are included in the archived repository.

\begin{table}[!t]
    \renewcommand{\arraystretch}{1.1}
    \begin{center}
        \caption{System Model and Controller Parameters}
        \label{tab:parameters}
        \begin{tabular}{l l}
            \hline
            \textit{General} & \\
            \hline
            Nominal voltage (line-to-line, rms) & $400$ V\\
            Nominal frequency & $50$ Hz\\
            Controller sampling frequency & $10$ kHz\\
            Excitation-voltage magnitude & $0.01\ldots0.05$ p.u.\\
            \hline
            \textit{DO-GFM control} & \\
            \hline
            Nominal current (rms) & $18$ A\\
            Nominal power & $12.5$ kVA\\
            Filter inductance & $0.15$ p.u.\\
            Active resistance $\Ra$ & $0.2$ p.u.\\
            Voltage-control gain $\kv$ & $1$\\
            Observer gain $\alphao$ & $2\pi\cdot50$ rad/s\\
            \hline
            \textit{GFL control} & \\
            \hline
            Nominal current (rms) & $14.5$ A\\
            Nominal power & $10$ kVA\\
            Filter inductance & $0.2$ p.u.\\
            Reference-tracking bandwidth $\alphac$ & $2\pi\cdot400$ rad/s \\
            Integral-action bandwidth $\alphai$ & $2\pi\cdot400$ rad/s \\
            PLL frequency-tracking bandwidth $\alphapll$ & $2\pi\cdot20$ rad/s\\
            \hline
        \end{tabular}
    \end{center}
\end{table}

\subsection{Grid-Forming Control}
\label{sec:do-gfm}

\subsubsection{Identified Output Admittance}
\begin{figure}[!t]
    \centering
    \includegraphics[width=\linewidth]{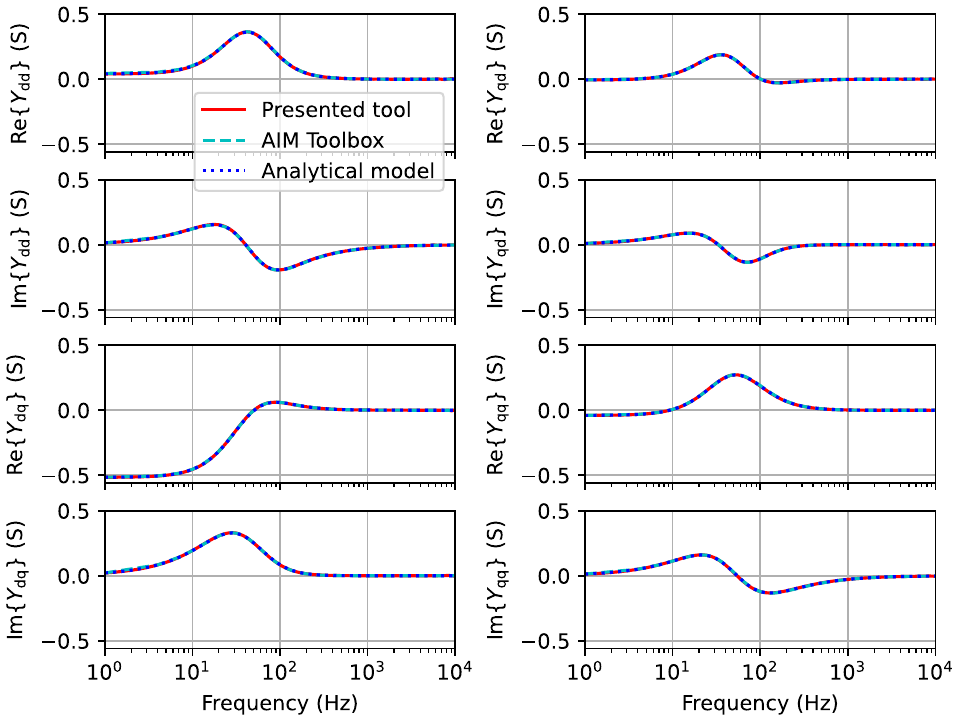}
    \caption{Comparison of numerically identified and analytically calculated output admittance using disturbance-observer-based grid-forming control.}
    \label{fig:results_do-gfm}
\end{figure}

In this section, converter output admittance is obtained for the DO-GFM control method. Fig.~\ref{fig:results_do-gfm} shows the comparison of the results obtained using the presented numerical tool, the commercial AIM Toolbox for PSCAD, and the analytical model from Section~\ref{sec:do-gfm_admittance}. Unlike in \textit{motulator}, in PSCAD the whole simulation model is implemented in continuous time including the control system. The operating point is set to $\pgo=0.5$ p.u. and $\ucmago=1$ p.u. Furthermore, the converter is connected to a weak grid with $\Lg=0.74$ p.u., corresponding to $\qgo=0.078$ p.u. The total inductance estimate is $\hatLt = 0.15$ p.u. It can be seen that the results from the presented numerical tool match very well with both the commercial tool and the analytical model.

\subsubsection{Computational Requirements}
The computational requirements of the presented identification tool are evaluated by comparing its execution time with that of AIM Toolbox. The test case shown in Fig.~\ref{fig:results_do-gfm} is used for the evaluation. The tested versions were PSCAD 5.0.1 and AIM Toolbox 5.1.1. The results were obtained on a laptop running Windows 11 with an Intel Core Ultra 5 125U processor and 32 GB of memory. To account for variations in the execution times due to background processes, two runs were performed for each tool, alternating between the two tools. Multiprocessing was enabled for the presented tool, and the broadband excitation option was used for AIM Toolbox.

The average execution times over the two runs were 2 min 54 s for the presented tool and 16 min 9 s for AIM Toolbox. The Python-based tool performs the same measurement substantially faster than AIM Toolbox. The difference is mainly due to the repeated compilation of the PSCAD model during the identification process. With one hundred measurement frequencies and broadband excitation enabled, the PSCAD model is compiled 40 times during the identification process. For high excitation frequencies and thus short simulation lengths, the compilation takes substantially longer than the actual simulation. This compilation overhead is avoided in the Python-based implementation, where the execution time is mainly determined by the simulation length, assuming that other parameters such as simulation time step remain constant.

\subsubsection{Passivity Index}

The passivity index is considered here as an application example, demonstrating how the identified output admittance can be further used for stability analysis. In the circuit theory sense, a passive system comprises only passive components, and thus can only store or dissipate but not produce energy. A precondition for a system being passive is that it is stable. If the converter output admittance is passive, stability is then guaranteed when the converter is connected in parallel with other passive devices~\cite{harnefors2017}.

For a single-input single-output system, passivity is defined as the output admittance having a nonnegative real part across all frequencies. For the MIMO converter system, the passivity condition is~\cite{harnefors2007a}
\begin{align}\label{eq:mimo_passivity}
    \mathrm{min}\left(\lambda\left[\Yc (\jj\omega) + \Yc ^\mathrm{H}(\jj\omega)\right]\right) > 0,\ \forall \omega>0
\end{align}
where $\mathrm{min}\left(\lambda\left[\cdot\right]\right)$ denotes the minimum eigenvalue of a matrix and the superscript $\mathrm{H}$ denotes the Hermitian (complex conjugate) transpose. For a passive grid impedance, a sufficient but not necessary condition for stability of the closed-loop system is that the converter output admittance is also passive.

\begin{figure}[!t]
    \centering
    \includegraphics[width=0.6\linewidth]{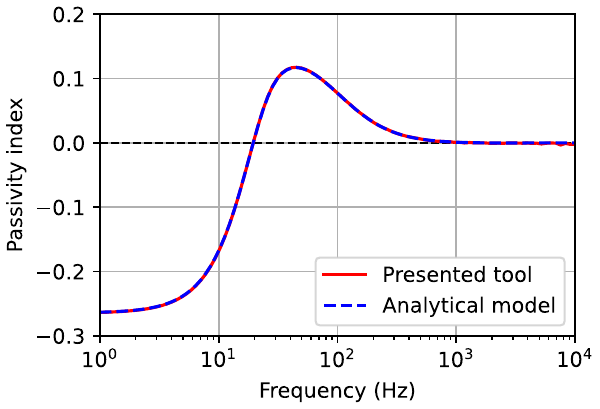}
    \caption{Input feedforward passivity index for the disturbance-observer-based grid-forming control method.}
    \label{fig:passivity_do-gfm}
\end{figure}

From~\eqref{eq:mimo_passivity}, the input feedforward passivity index can be derived as~\cite{chen2024}
\begin{align}
    \nuF \left[\Yc (s),\omega\right] = \frac{1}{2}\ \mathrm{min} \left(\lambda\left[\Yc (\jj\omega) + \Yc ^\mathrm{H}(\jj\omega)\right]\right).
\end{align}
Fig.~\ref{fig:passivity_do-gfm} shows this index for the DO-GFM control method using the presented tool and the analytical model. The results match very well.

\subsection{Grid-Following Control}
\begin{figure}[!t]
    \centering
    \includegraphics[width=\linewidth]{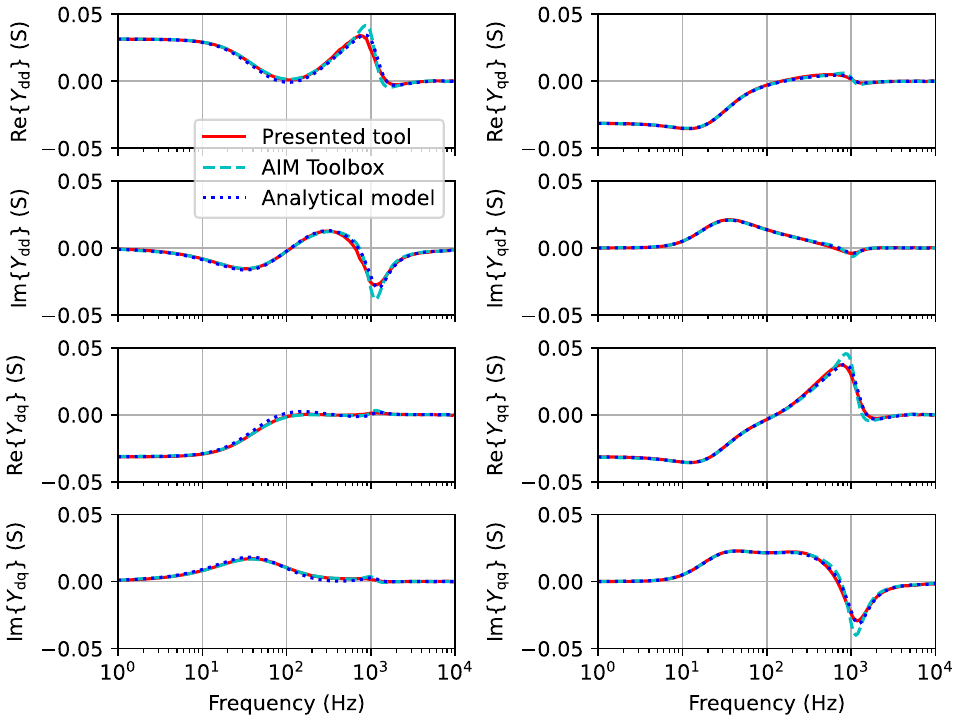}
    \caption{Comparison of numerically identified and analytically calculated output admittance using grid-following control.}
    \label{fig:results_gfl}
\end{figure}

In this section, the converter output admittance is obtained for the GFL control method. Fig.~\ref{fig:results_gfl} shows the comparison of the results from the presented numerical tool, the commercial AIM Toolbox for PSCAD, and the analytical model from Section~\ref{sec:gfl_admittance}. Again, in PSCAD, the control system is implemented in continuous time. The operating-point active and reactive powers at the PCC are $\pgo=0.5$~p.u. and $\qgo=0.5$~p.u., respectively. Furthermore, $\Lg=0$ and $\hatLt=0.2$~p.u. It can be seen that the results from the presented numerical tool and the analytical model match very well.

\section{Conclusion}
\label{sec:conclusion}

Frequency-domain impedance-based methods have become important tools in the stability analysis of grid converters. This paper presented a new Python-based grid converter output-admittance identification tool developed on the open-source \textit{motulator} platform. The results obtained with the presented tool were verified against those from a commercial PSCAD-based identification tool and analytical models. 

In future research, the presented tool could be further developed to include broadband excitation signals for faster identification. A parameter-sweep feature could also be added, allowing for easier identification using various operating points or control gains.


\end{document}